\documentclass[aps, prr, twocolumn,amssymb,showpacs,superscriptaddress,nofootinbib,longbibliography,10pt]{revtex4-2}
\usepackage[english]{babel}
\usepackage[utf8]{inputenc}
\usepackage[T1]{fontenc}
\usepackage{MnSymbol}
\usepackage{amsmath}
\usepackage{mathtools}
\usepackage{xcolor}
\usepackage{amsfonts}
\usepackage{comment}
\usepackage{braket}
\usepackage{float}
\usepackage{subfigure}
\usepackage[colorlinks=true,linktoc=page,linkcolor=blue,citecolor=blue,urlcolor=blue]{hyperref}
\usepackage{multirow}
\usepackage{enumerate}
\usepackage{graphicx}
\usepackage{bm}
\usepackage{units}
\usepackage{sidecap,tikz}
\usepackage[normalem]{ulem}

\mathchardef\mhyphen="2D 

\definecolor{lime}{HTML}{A6CE39}
\DeclareRobustCommand{\orcidicon}{\hspace{-1.0mm}
	\begin{tikzpicture}
		\draw[lime, fill=lime] (0.0,0.0) 
		circle [radius=0.15] 
		node[white] {{\fontfamily{qag}\selectfont \tiny \,ID}};
		\draw[white, fill=white] (-0.0525,0.095) 
		circle [radius=0.007];
	\end{tikzpicture}
	\hspace{-3.0mm}
}
\newcommand{\orcidVAM}{\href{https://orcid.org/0000-}{\orcidicon}}

\allowdisplaybreaks

\begin{document}

\title{Variational M-Partite Geometric Entanglement Algorithm}  

\author{Vahid Azimi-Mousolou\orcidVAM}
\email{vahid.azimi-mousolou@physics.uu.se}
\affiliation{Department of Physics and Astronomy, Uppsala University, Box 516, 
SE-751 37 Uppsala, Sweden}

\author{Prashant Singh}
\affiliation{Science for Life Laboratory, Department of Information Technology, Uppsala University, Box 337, SE-751 05 Uppsala, Sweden}

\begin{abstract}
Variational quantum algorithms have emerged as a powerful tool for harnessing the potential of near-term quantum devices to address complex challenges across quantum science and technology. Yet, the robust and scalable quantification of entanglement in many-body quantum systems remains a significant challenge, crucial for both advancing theoretical understanding and enabling practical applications. In this work, we propose a variational quantum algorithm to evaluate the $M$-partite geometric entanglement across arbitrary partitions of an $N$-qubit system into $M$ parties. By constructing tailored variational ansatz circuits for both single- and multi-qubit parties, we optimize the overlap between a target quantum state and an $M$-partite variational separable state. This method provides a flexible and scalable approach for characterizing arbitrary $M$-partite entanglement in complex quantum systems of a given dimension.  
The accuracy of the proposed method is assessed by reproducing known analytical results. We further demonstrate its capability to evaluate entanglement among $M$ parties for any given conventional or unconventional partitions of one- and two-dimensional spin systems, both near and at a quantum critical point. Our results establish the versatility of the variational approach in capturing different types of entanglement in various quantum systems, surpassing the capabilities of existing methods.  
Our approach offers a powerful methodology for advancing research in quantum information science, condensed matter physics, and quantum field theory. Additionally, we discuss its advantages, highlighting its adaptability to diverse system architectures in the context of near-term quantum devices.  
\end{abstract}

\maketitle

\section{Introduction} \label{Sec:introduction}
Entanglement is a cornerstone of quantum mechanics and plays a critical role in diverse areas ranging from quantum information processing to
condensed matter physics and quantum field theory \cite{Horodecki2009}. 
While bipartite entanglement measures, such as concurrence and negativity, are well-established, 
quantifying entanglement, especially in multipartite systems, remains a challenging task due to the exponential growth of the Hilbert space and the complexity of separability criteria \cite{Horodecki2009, Plenio2007}. 

Existing methods, such as the density matrix renormalization group (DMRG) \cite{Eisert2010, Schollwock2011} and tensor network techniques, including matrix product states (MPS), projected entangled pair states (PEPS) \cite{Verstraete2008}, and tree tensor networks (TTN) \cite{Pizorn2013}, have provided significant insights into ground state properties and entanglement structures, particularly in one-dimensional systems. While DMRG, which is closely related to MPS, is highly efficient for 1D gapped systems, it struggles with two- and higher-dimensional models due to an exponential growth in computational cost. MPS itself faces similar limitations, particularly in capturing long-range entanglement and critical systems. PEPS generalizes MPS to higher dimensions but suffers from computationally expensive tensor contractions and optimization challenges. TTN, while effective for hierarchical structures, lacks translation invariance and is less suited for capturing entanglement in complex geometries. Stochastic approaches, like quantum Monte Carlo methods, including variational Monte Carlo (VMC) \cite{Becca2017}, on the other hand, offer greater flexibility in wavefunction design and can handle higher-dimensional systems. While VMC has primarily been focused on energy minimization, recent studies have demonstrated its potential for evaluating bipartite and multipartite entanglement \cite{Ferris2012, Deng2017, Harney2020}. Despite these advances, the available methods still face significant challenges in quantifying multipartite entanglement and characterizing entanglement structures in high-dimensional, complex geometries. Therefore, there is a strong need for efficient and accurate approaches to evaluate various types of entanglement, particularly beyond bipartite entanglement.

Here, we present a quantum variational $M$-partite geometric entanglement (VMGE) algorithm, which is generic and dresses the current challenges to a great extent. Our method 
considers $N$-qubit system partitioned into an arbitrary $M$ parties, defining $M$-partite geometric entanglement (GE) through the minimal distance between the target state and the set of
parametrized variational M-partite separable states. We construct variational quantum circuits tailored to each party, optimizing over universal parametrized single-qubit rotations and multi-qubit entangling gates with variational entanglement power to maximize overlap with the target state. The optimization is driven by a synergy between the non-gradient variational Monte Carlo method and gradient-based Broyden–Fletcher–Goldfarb–Shanno (BFGS) method \cite{Nocedal1999}. The VMGE algorithm provides a hybrid quantum-classical framework suitable for near-term quantum devices.

We demonstrate the efficacy of our approach through detailed mathematical formulation and circuit design. Our results highlight the algorithm's scalability, flexibility, making it a promising universal tool for multipartite entanglement quantification in complex quantum systems. This work opens new avenues for exploring multipartite entanglement in both theoretical and experimental settings, with potential applications in quantum computing, quantum simulation, and beyond.

\section{Variational M-Partite Geometric Entanglement Algorithm} \label{Sec:Methods}
\subsection{Variational quantum algorithm}
For an $N$-qubit system partitioned into $M$ parties, let us assume $\mathcal{S}_M$ is the set of all M-partite separable states, i.e., states that can be written as
\begin{eqnarray}
  |\phi\rangle = \ket{\phi_1} \otimes \ket{\phi_2}\otimes \cdots \otimes \ket{\phi_M},
\end{eqnarray}
 where each $|\phi_i\rangle$ is a state involving $m_i$ qubits, with $\sum_{i=1}^M m_i = N$. 
For a given $N$-qubit target state, $|\psi\rangle$, the M-partite GE is defined as \cite{Wei2003}
\begin{eqnarray}
E^{(M)}(|\psi\rangle) =1 -\max_{|\phi\rangle \in \mathcal{S}_M} |\langle \phi | \psi \rangle|^2.
\end{eqnarray}

\begin{figure}[h]
\begin{center}
\includegraphics[width=85mm]{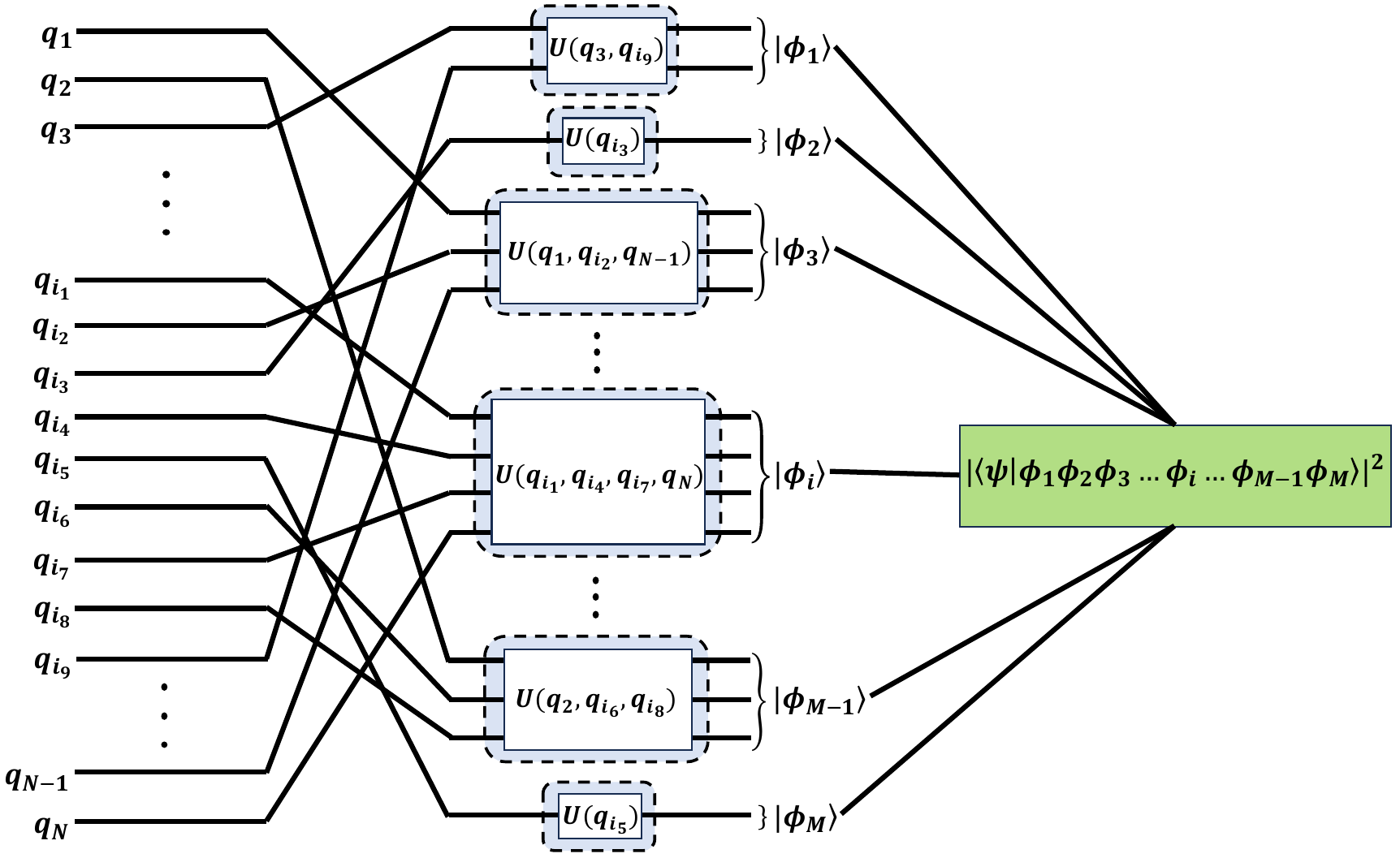}
\end{center}
\caption{(Color online) Variational M-partite geometric entanglement (VMGE) algorithm. For an $N$-qubit system, the algorithm involves $M$ disjoint variational circuits corresponding to a given partition of the system into $M$ parties. 
These circuits generate a general $M$-partite separable state, which is then compared with a given $N$-qubit target state $ \ket{\psi} $. The geometric entanglement is evaluated through an optimization over the separable state space.
}
\label{fig1}
\end{figure}

\begin{figure}[h]
\begin{center}
\includegraphics[width=75mm]{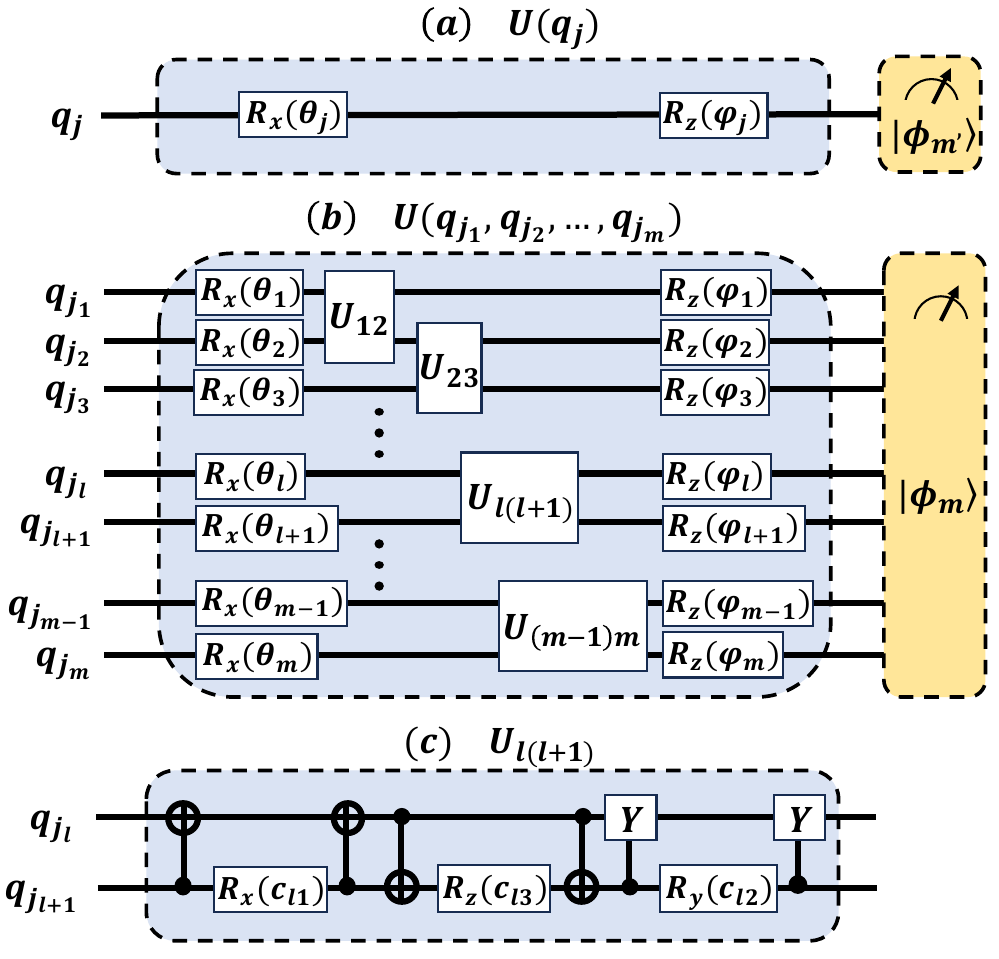}
\end{center}
\caption{(Color online) Variational quantum circuits: variational single-qubit ($a$) and multi-qubit ($b$) circuits used in the VMGE algorithm, as illustrated in Fig.~\ref{fig1}. The algorithm employs universal single-qubit circuits combined with a two-qubit entangling circuits ($c$) with variational entangling power.}
 \label{fig2}
   \end{figure}

Fig. \ref{fig1} illustrates a variational quantum algorithm to evaluate the overlap of between the $N$-qubit target state and a variational ansatz for $M$-Partite Separable States.
For each party involving $m$ qubits, the variational ansatz is constructed through a variational circuit, $U(q_{i_1}, q_{i_2}, \dots, q_{i_m})$, shown in Fig \ref{fig2}. 
The circuit is described as follows
\begin{itemize}
\item [i)] Single-qubit party, $m=1$: As shown in Fig \ref{fig2} ($a$) the variational circuit simplifies to
\begin{eqnarray}
U(q_j) =R_z(\varphi_j) R_x(\theta_j) 
\end{eqnarray}
\item [ii)] Multi-qubit party, $m>1$: In this case 
\begin{eqnarray}
U(q_{i_1}, q_{i_2}, \dots, q_{i_m})=\bigotimes_{l=1}^{m}R_z(\varphi_{l})\bigotimes_{l=1}^{m-1}U_{l(l+1)} \bigotimes_{l=1}^{m}R_x(\theta_{l})\nonumber\\
\end{eqnarray}
specifies the variational circuit in Fig \ref{fig2} ($b$), where $U_{l(l+1)}$ is a two-qubit entangling gate on nearest-neighbor qubit pairs.
We assume a general nonlocal two-qubit operation
\begin{eqnarray}
U_{l(l+1)} = \exp\left[-\frac{i}{2}\sum_{i=1}^{3}c_{li}\sigma_{i}^{l}\sigma_{i}^{l+1}\right]
\end{eqnarray}
where $c_{l1}, c_{l2}$, and $c_{l3}$ are variational parameters controlling the entanglement power of the circuit up to arbitrary value. 
The real vector $(c1, c2, c3)$ is a geometric coordinate on a 3-Torus, which represents local invariants restricted to the tetrahedral Weyl chamber \cite{Zhang2003}.
Fig \ref{fig2} ($c$) is the circuit representation of $U_{l(l+1)}$.
 \end{itemize}  

By initializing all $N$ qubits in the $|0\rangle$ state, the overall M-Partite variational separable state is
\begin{eqnarray}
|\phi(\vec{\theta}, \vec{\varphi}, \vec{c})\rangle = \bigotimes_{j=1}^M U(q_{i_1}, q_{i_2}, \dots, q_{i_{m_j}}) |0\rangle^{\otimes N}.
\end{eqnarray}
This lead to the objective function for measureing the overlap between the target state $|\psi\rangle$ and the variational state $|\phi\rangle$ as
\begin{eqnarray}
     \mathcal{L}(\vec{\theta}, \vec{\varphi}, \vec{c}) = |\langle \phi(\vec{\theta}, \vec{\varphi}, \vec{c}) | \psi \rangle|^2.
\end{eqnarray}
Using a classical optimizer the proposed variational quantum algorithm allows to compute the M-partite GE through the optimization 
\begin{eqnarray}
     E^{(M)}(|\psi\rangle) &=& 1-\max\mathcal{L}(\vec{\theta}, \vec{\varphi}, \vec{c})\nonumber\\
     &=&\min\left(1-\mathcal{L}(\vec{\theta}, \vec{\varphi}, \vec{c})\right),
\end{eqnarray}
where maximize and minimum are taken over the variational parameters $\vec{\theta}, \vec{\varphi}, \vec{c}$.

\subsection{Optimization procedure.}
We solve the optimization problem formulated above using a two-stage hybrid optimization process. In the first stage, a Sobol sequence \cite{joe2008} of $n_{init}=2000$ samples is used to perform a 'global' Monte-Carlo search within the interval $[0, 2\pi]$ for each parameter to identify the parameter combination $\vec{\theta}^*, \vec{\varphi}^*, \vec{c}^*$ offering the best overlap. This combination is then used as the starting point in the second stage composed of optimization using the quasi-Newton Broyden–Fletcher–Goldfarb–Shanno (BFGS) method \cite{Nocedal1999}.

\section{Benchmarking and Applications} \label{Sec:Results}
\subsection{Validation of VMGE algorithm} 
To validate our scheme, we evaluate different types of GE for quantum states with known analytical solutions
\begin{eqnarray}
\ket{\psi(p)} &=& \sqrt{p} \ket{00} + \sqrt{1-p} \ket{11}\nonumber\\
\ket{W(p, \phi)} &=& \sqrt{p} \ket{W} + \sqrt{1-p} e^{i\phi} \ket{\tilde{W}}\nonumber\\
\ket{GW(p, \phi)} &=& \sqrt{p} \ket{GHZ} + \sqrt{1-p} e^{i\phi} \ket{W}\nonumber\\
\ket{BB(p)} &=& \sqrt{f(p)}[\ket{0000}+ \ket{1111}]/\sqrt{2} \nonumber\\
&&+  \sqrt{1-f(p)}[\ket{0101}+\ket{1010}]\sqrt{2}, \nonumber\\
\label{States}
\end{eqnarray}
where $f(p)=\frac{1+ 2\sqrt{p(1 - p)}}{2}$,
$\ket{W} = \frac{1}{\sqrt{3}} \Big( \ket{001} + \ket{010} + \ket{100} \Big)$,
$\ket{\tilde{W}}= \frac{1}{\sqrt{3}} \Big( \ket{011}+ \ket{101} + \ket{110} \Big)$,
and $\ket{GHZ}=\frac{1}{\sqrt{2}}(\ket{000}+\ket{111}$.
Considering $|\psi_{xy}^{\pm}\rangle=\frac{1}{\sqrt{2}}( |00\rangle\pm|11\rangle)$ as the Bell states corresponding to pair of qubits x and y, we have
\begin{eqnarray}
\ket{BB(p)} = \sqrt{p}\ket{\psi_{13}^{+}}\ket{\psi_{24}^{+}}+ \sqrt{1 - p}\ket{\psi_{13}^{-}}\ket{\psi_{24}^{-}}.
\end{eqnarray}

\begin{figure}[h]
\begin{center}
\includegraphics[width=70mm]{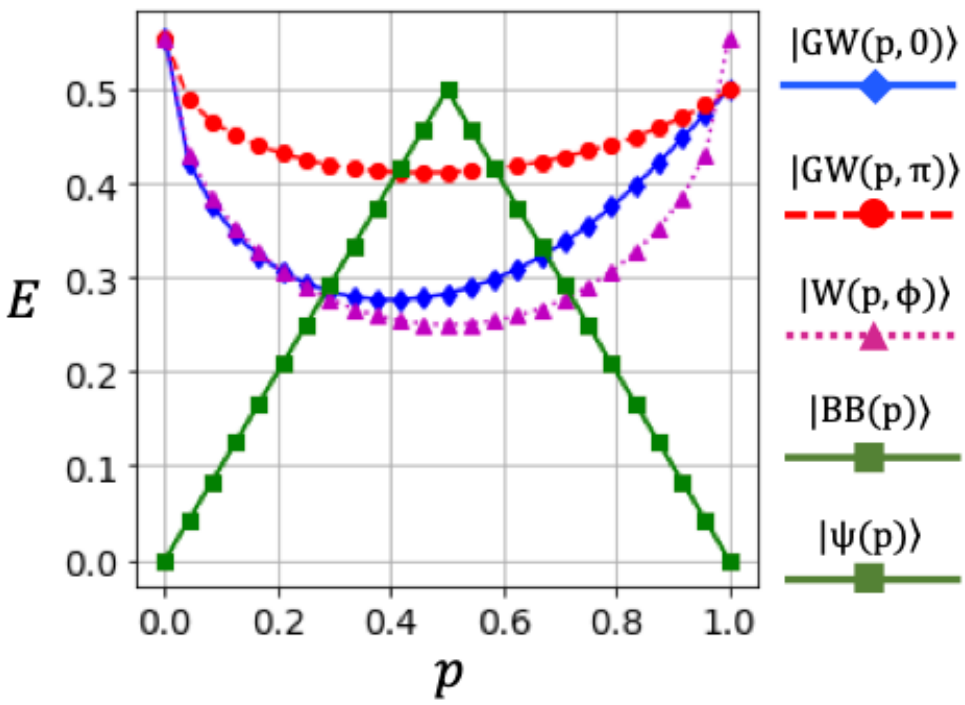}
\end{center}
\caption{(Color online) Validation of the VMGE algorithm in determining exact GE values. The VMGE algorithm is used to evaluate the bipartite GE in $ \ket{\psi(p)} $, the tripartite GE in $ \ket{W(p, \phi)} $, $ \ket{GW(p, 0)} $, and $ \ket{GW(p, \pi)} $, as well as the bipartite GE between subsystems $ S_1 = \{\text{qubit 1, qubit 3}\} $ and $ S_2 = \{\text{qubit 2, qubit 4}\} $ in the four-qubit state $ \ket{BB(p)} $, all shown as functions of $p$. The resulting plots coincide precisely with the exact GE values for these states reported in Ref.~\cite{Wei2003}.}
 \label{fig11}
 \end{figure}

The exact GEs for the states in Eq.\eqref{States} were studied in Ref.~\cite{Wei2003}. In Fig.\ref{fig11}, we demonstrate the accuracy of the VMGE in precisely reproducing the exact solutions obtained in~\cite{Wei2003}.
The figure illustrates the bipartite GE in $ \ket{\psi(p)} $, the tripartite GE in $ \ket{W(p, \phi)} $, $ \ket{GW(p, 0)} $, and $ \ket{GW(p, \pi)} $, 
as well as the bipartite GE between subsystems $ S_1 = \{\text{qubit 1, qubit 3}\} $ and $ S_2 = \{\text{qubit 2, qubit 4}\} $ in the four-qubit state $ \ket{BB(p)} $.

\subsection{Applications to critical phenomena in many-body spin models} 
We consider various unconventional partitions to compute the $M$-partite geometric entanglement (GE) of the ground state for two prototypical spin models,
 the transverse-field XY model and the anisotropic antiferromagnetic Heisenberg (XXZ) model. Their Hamiltonians are given by
\begin{eqnarray}
H_{\text{XY}} &=& -J \sum_{\langle i,j \rangle} \left( \frac{1+r}{2} \sigma_i^x \sigma_j^x + \frac{1-r}{2} \sigma_i^y \sigma_j^y \right) - h \sum_{i} \sigma_i^z, \nonumber \\
H_{\text{XXZ}} &=& J \sum_{\langle ij \rangle} \left( \sigma_i^x \sigma_{j}^x + \sigma_i^y \sigma_{j}^y + \Delta \sigma_i^z \sigma_{j}^z \right) - h \sum_{i} \sigma_i^z, \nonumber
\end{eqnarray}
where $\sigma_i^x$, $\sigma_i^y$, and $\sigma_i^z$ are Pauli matrices. 
These models involve various spin systems subjected to an external magnetic field and are known to describe a broad class of critical quantum systems. The 
$H_{\text{XXZ}}$ Hamiltonian with $\Delta=1$ corresponds to an isotropic antiferromagnetic Heisenberg interaction, while the 
$H_{\text{XY}}$ Hamiltonian reduces to the transverse-field Ising model when $r=1$. In the special case where 
$\Delta=r=0$, both Hamiltonians reduce to the XX model. We analyze these models in both one-dimensional (1D) and two-dimensional (2D) systems under periodic boundary conditions.

The study of ground-state entanglement in spin models has long been a subject of great interest across various branches of physics, particularly in quantum information science and condensed matter physics. Extensive qualitative analyses have been conducted on ground-state entanglement in one-dimensional spin chains, with a particular focus on the XY model, which is exactly solvable within the spinless fermionic representation \cite{Sachdev1999, Osterloh2002, Osborne2002a, Vidal2003, Wei2005, Orus2008, Orus2008a, Orus2011, Orus2011a, Son2011, Azimi-Mousolou2013}. Most existing studies have concentrated on two-spin entanglement, global entanglement, or bipartite entanglement for specific partitions of the chain. However, a more comprehensive understanding of entanglement structure beyond these cases remains an open challenge, motivating further exploration of multipartite and subsystem entanglement in quantum spin models. As discussed below, our proposed VMGE method allows to overcome these limitations and analyze multipartite geometric entanglement for arbitrary partitions and dimensions.

\begin{figure}[h]
\begin{center}
\includegraphics[width=85mm]{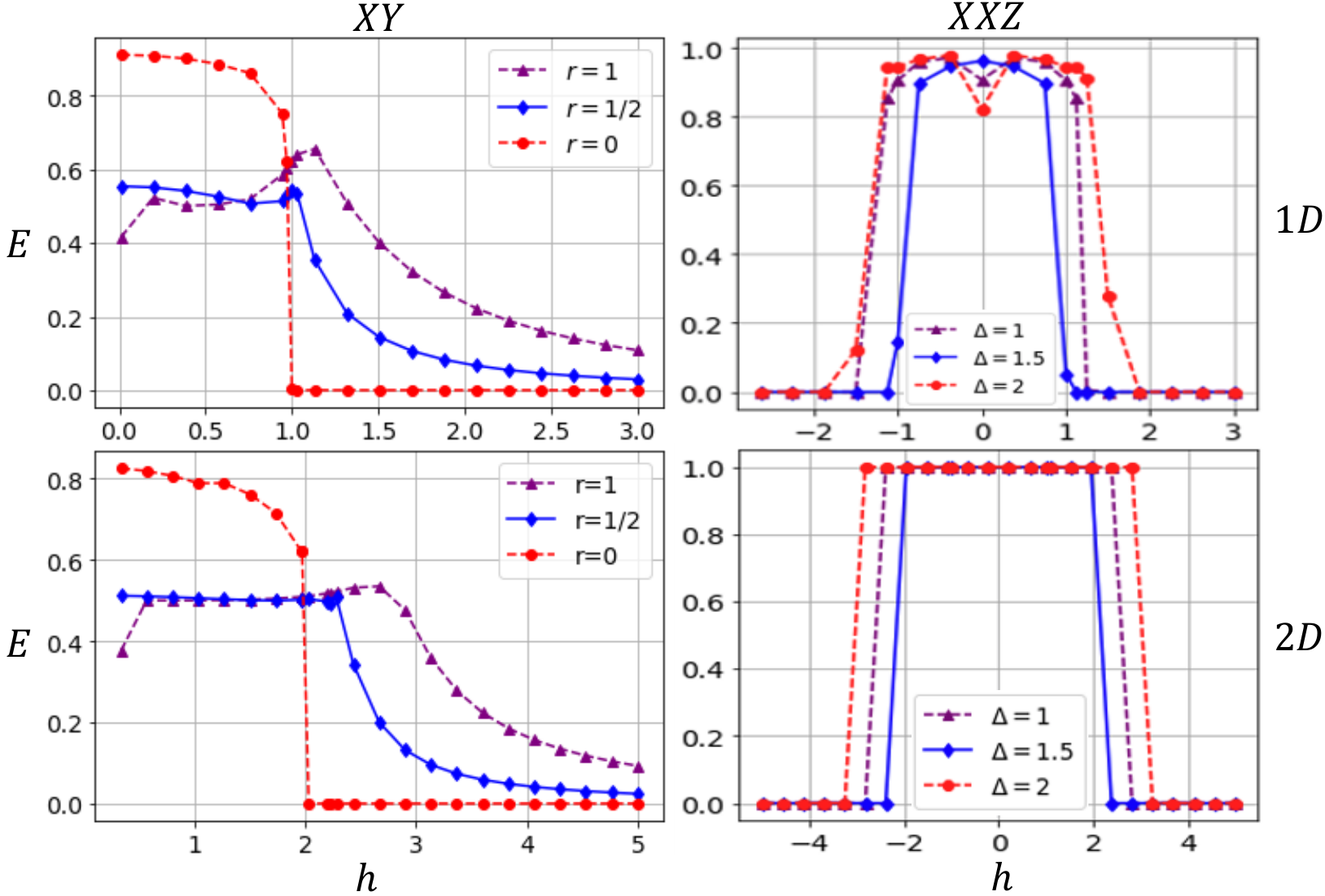}
\end{center}
\caption{(Color online) Results of the VMGE algorithm for evaluating 16-partite global GE in many-spin systems. Ground-state global GE for the XY (left column) and XXZ (right column) models in 1D (top row) and 2D (bottom row), obtained using the VMGE algorithm.  We consider a 16-spin ring for the 1D system and a $4\times4$ square lattice for the 2D system. The results demonstrate that the VMGE algorithm successfully handles critical regions and accurately captures quantum phase transitions in all cases. We assume $J=1$ for the XY model and $J=0.25$ for XXZ model.}
\label{1D16t}
\end{figure}

In Fig.~\ref{1D16t}, we analyze the ground-state global GE for the XY and XXZ models in 1D spin ring and 2D square lattice architectures. As illustrated in the figure, the method effectively handles critical systems and accurately detects quantum phase transitions at the critical fields, $h_c \approx J$ for the 1D XY model, $h_c \approx 3J$ for the 2D XY model, and $h_c \approx J\mathcal{Z}(1+\Delta)$ for the 1D and 2D XXZ models. Here, $\mathcal{Z}$ is the coordination number of the lattice, i.e., the number of nearest neighbors for each spin site, with $\mathcal{Z} = 2$ for a 1D spin chain and $\mathcal{Z} = 4$ for a 2D square lattice.

\begin{figure*}[t]
\begin{center}
\includegraphics[width=145mm]{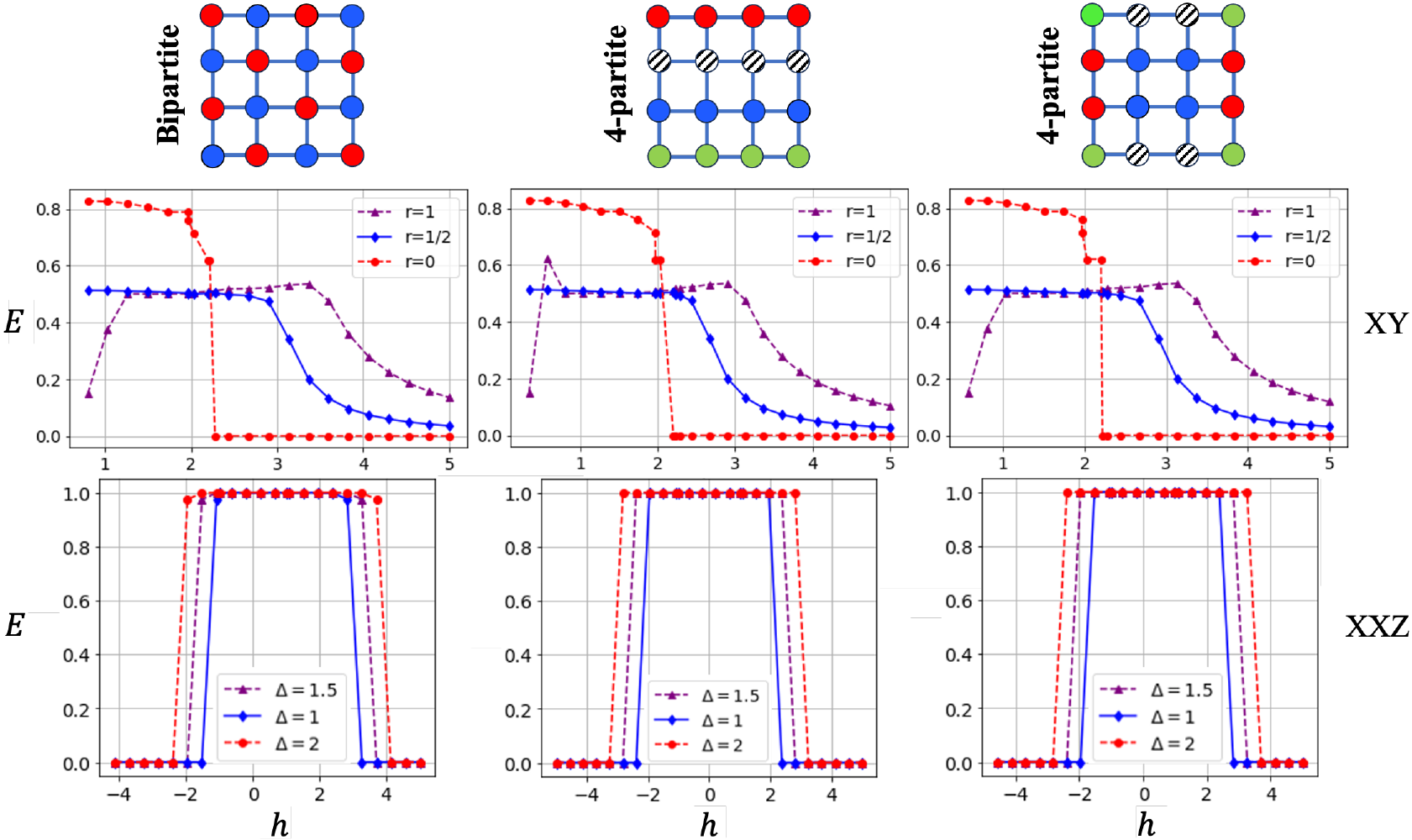}
\end{center}
\caption{(Color online) Results of the VMGE algorithm for evaluating GEs associated with unconventional partitions in many-spin systems. Ground-state 2- and 4-partite GEs for the XY (middle row) and XXZ (bottom row) models in 2D $4\times4$ square lattice, obtained using the VMGE algorithm. Each column correspond to GE associated to an unconventional partitions indicated by a colored square lattice (top row). The results demonstrate that the VMGE algorithm successfully handles critical regions and accurately captures quantum phase transitions in all cases. We assume $J=1$ for the XY model and $J=0.25$ for XXZ model.}
\label{2D16Mp}
\end{figure*}

 We further analyze the ground-state bipartite and four-partite GEs for the XY and XXZ models in a $4 \times 4$ 2D square lattice architecture in Fig.~\ref{2D16Mp}. We consider unconventional partitions, for which existing methods such as DMRG, MPS, PEPS, or TTN struggle to evaluate quantum entanglement. The figure indicates that the VMGE algorithm effectively evaluates entanglement in such unconventional scenarios at critical regions and accurately detects quantum phase transitions.  

Both Figs.~\ref{1D16t} and \ref{2D16Mp} highlight the flexibility of the VMGE algorithm in characterizing arbitrary $M$-partite entanglement in complex critical quantum systems of varying dimensions. This flexibility, combined with the accuracy of the VMGE algorithm, as confirmed by Fig.~\ref{fig11}, establishes significant advantages over existing methods.  Another key advantage of the VMGE algorithm is its adaptivity, which allows symmetries inherent to a given target state or quantum system to be incorporated into the quantum circuit design. Taking these symmetries into account leads to a substantial reduction in the dimensionality of the optimization problem, thereby enhancing both the efficiency and scalability of the algorithm. For instance, when evaluating the global GE in a spin-$\frac{1}{2}$ quantum system with translational invariance, such as the spin ring and square lattice architectures considered above, incorporating the symmetry of the lattice structures reduces the problem to a two-variable real function optimization, regardless of the number of spins. This demonstrates that the VMGE algorithm naturally and, in principle, scales to an arbitrary number of spins in translationally invariant spin models, provided that the target state is specified.

\begin{figure}[h]
\begin{center}
\includegraphics[width=85mm]{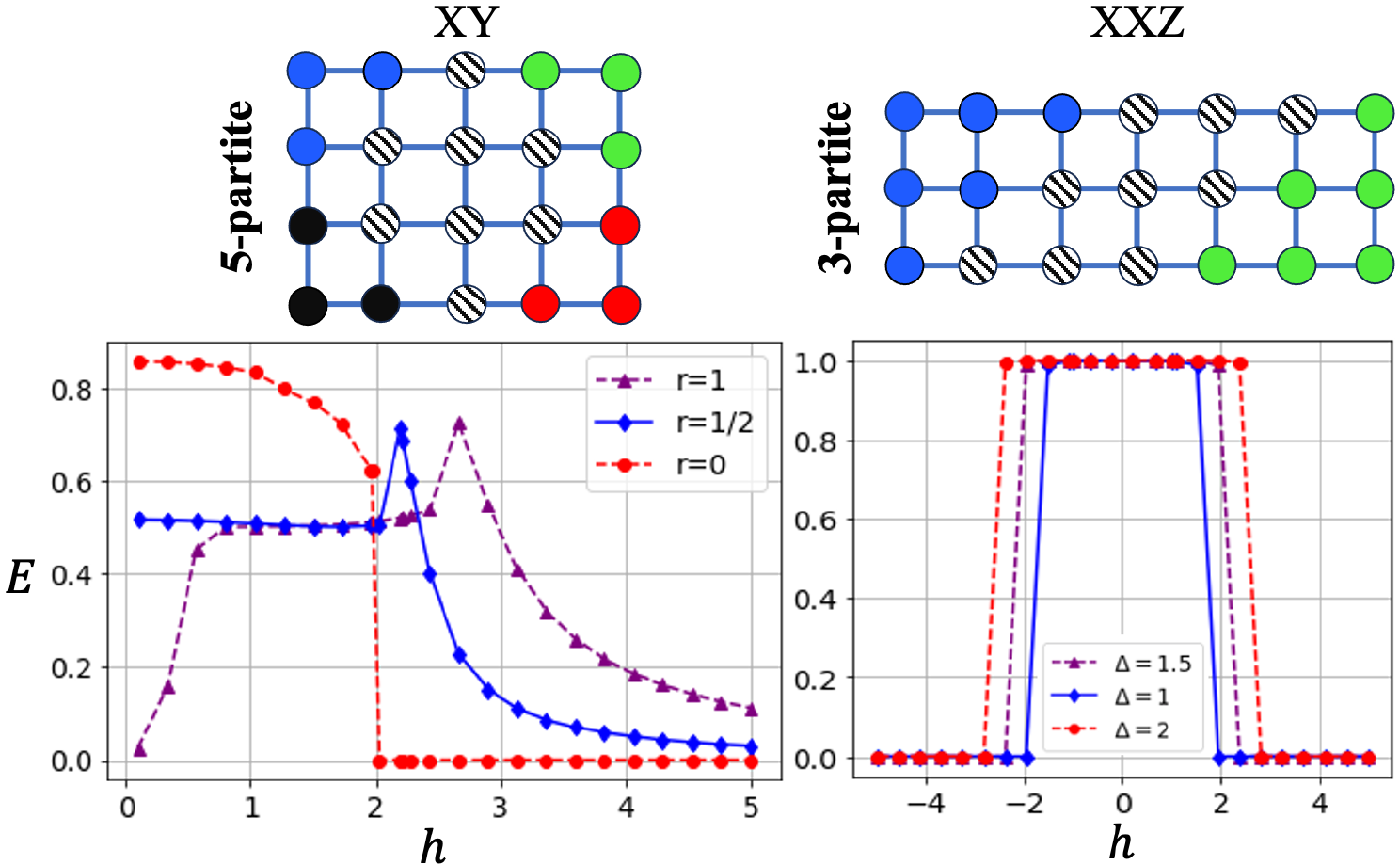}
\end{center}
\caption{(Color online) Results of the VMGE algorithm for evaluating unconventional 5-partite and 3-partite GEs in many-spin systems. Ground-state 5-partite GE for the XY model (left column) within a 2D 20-spin cluster, and 3-partite GE for the XXZ model (right column) within a 2D 21-spin cluster, obtained using the VMGE algorithm. We assume $J = 1$ for the XY model and $J = 0.25$ for the XXZ model. The VMGE algorithm successfully handles critical regions and accurately captures quantum phase transitions in both cases.
}
\label{fig4}
\end{figure}

While in Figs.\ref{1D16t} and \ref{2D16Mp} we focus on 16-spin clusters, since the geometric entanglements have already converged and the main results and features are clearly demonstrated, we have also explored larger clusters of 20, 21, and 24 spins using our limited local hardware. These tests, employing various unconventional partitions, confirm that the VMGE algorithm is consistently capable of handling larger system sizes. Figure~\ref{fig4} illustrates examples of unconventional 5-partite and 3-partite geometric entanglements within 20-spin and 21-spin clusters, respectively.

Despite specializing our discussion and examples to qubit (spin-$\frac{1}{2}$) systems for the sake of concreteness, it is important to note that the VMGE algorithm can be routinely adapted to address GEs in quantum systems involving bosons, fermions, or similar particles. Notably, integrating the VMGE algorithm with widely used quantum computing frameworks such as Pennylane and Qiskit is straightforward, making it well-suited for implementation on near-term quantum devices. Furthermore, when combined with techniques such as variational quantum states based on artificial neural networks (NQS) \cite{Carleo2017} for finding ground-state representations of quantum systems, it develops into a highly effective methodology for studying quantum phenomena in many-body quantum systems.

\section{Summary} \label{Sec:summary}
We present a variational quantum algorithm, VMGE, and demonstrate its capability to evaluate M-partite geometric entanglement across arbitrary partitions of an N-body quantum system into M parties. Despite its simplicity, the VMGE algorithm yields highly accurate results for both well-known target states and the ground-state geometric entanglement of prototypical spin models at quantum phase transitions. Furthermore, the VMGE algorithm provides an efficient and naturally scalable methodology that can be readily integrated into current and near-term quantum computing frameworks for studying quantum correlations in various quantum systems. We demonstrate that the VMGE algorithm is capable of evaluating types of quantum entanglement that existing methods struggle to handle.  Additionally, the VMGE algorithm can be seamlessly combined with existing techniques for finding ground-state representations of quantum systems. This enhances its adaptability and flexibility, potentially leading to a more powerful approach when integrated with other state-of-the-art methods. Such a methodology will be highly useful in quantum information science, condensed matter physics, and quantum field theory, paving the way for new insights and advancements.
\\


\section*{Acknowledgments} \label{Sec:Acknowledgments}
Vahid Azimi Mousolou (VAM) gratefully acknowledges the insightful discussions with Prof. Olle Eriksson and his generous support through research funding.

\end{document}